\documentclass[twocolumn,showpacs,preprintnumbers,amsmath,amssymb]{revtex4}
\usepackage{epsfig}
\usepackage{wasysym}
\usepackage{amssymb}

\begin{document}

\newcommand{\bvec}[1]{\mbox{\boldmath ${#1}$}}
\title{Electromagnetic production of $K\Sigma$ on the nucleon near threshold}
\author{T. Mart}
\affiliation{Departemen Fisika, FMIPA, Universitas Indonesia, Depok 16424, 
  Indonesia}
\date{\today}
\begin{abstract}
Photo- and electroproduction of $K\Sigma$ have been investigated 
near their production thresholds by using an effective Lagrangian 
approach. For this purpose, the background amplitude is constructed 
from suitable 
Feynman diagrams, whereas the resonance terms are calculated by
means of the Breit-Wigner form of multipoles.
Experimental data available in the proton channels ($K^+\Sigma^0$ and
$K^0\Sigma^+$)
with energies up to 50 MeV above the thresholds have been utilized 
to extract the unknown parameters. In these proton channels the 
calculated observables fit nicely the experimental data, whereas 
in the neutron channels ($K^+\Sigma^-$ and $K^0\Sigma^0$) the predicted
observables contain some uncertainties due to the the uncertainties 
in the values of helicity photon couplings. To this end, new
$K^0\Sigma^+$ photoproduction data are urgently required. 
The present analysis indicates the
validity of the $P_\Lambda = -\frac{1}{3} P_\Sigma$ relation derived
a long time ago.
In the electroproduction sector the present analysis confirms the smooth
transition between photoproduction and low $Q^2$ electroproduction
data.
The effect of new Crystall Ball data is shown to be mild at the
backward angles. 
It is also found that the electroproductions of $K^0\Sigma^+$ 
and $K^0\Sigma^0$  are practically 
not the suitable reactions for investigating
the $K^0$ charge form factor, since the effect is small.
\end{abstract}

\maketitle

\section{Introduction}
\label{sec:intro}
Meson photoproduction near its production threshold plays
a crucial role in improving our knowledge on the strong
interaction involving strangeness degree of freedom, 
since fewer parameters are involved and, therefore, only 
fewer uncertainties should be overcome at this kinematics. 
It is widely known that at energies where the
new and precise experimental data points mostly exist there
are more than 20 nucleon and 20 delta resonances listed
by Particle Data Group (PDG) \cite{olive} in the $s$-channel
of the reaction. There is also a number of hyperon and kaon resonances,
in the $u$- and $t$-channel, respectively, which should be taken
into account for a proper description of the kaon photoproduction process. 
Unfortunately, almost all of their coupling constants are 
hardly known and, therefore, must be treated as free parameters 
fitted to experimental
data.  As the energy increases the complexity of the problem also 
increases. For example, at a total center of momentum (c.m.) 
energy $W$ just above 2 GeV the
necessity to include hadronic form factors 
\cite{Haberzettl:1998eq,Davidson:2001rk,Mart:2013yka} and the phenomenon 
of Regge behavior seems to be inexorable 
\cite{Guidal:1997hy,Mart:2003yb,Corthals:2006nz,DeCruz:2012bv}. 
Clearly, such problems can be 
efficiently avoided by lowering the considered energy. 

In the previous works I have analyzed the electromagnetic 
production of $K^+\Lambda$ and $K^0\Lambda$ off a proton 
for energies up to 50 MeV above the thresholds  
\cite{mart_thr,mart_k0lambda}. For this purpose I made
use of an effective Lagrangian approach for the background 
terms and the Breit-Wigner form of multipoles for the 
resonance terms. Since the excitation energy 
was limited up to 50 MeV above 
threshold, only the $N(1650)S_{11}$ state could exist in the 
resonance terms. However, despite this substantial simplification, 
there have been very few studies
of kaon photoproduction devoted to the threshold region
\cite{mart_thr,mart_k0lambda,Steininger:1996xw,Cheoun:1996kn}. 
This is
understandable, since the corresponding theoretical analysis 
requires support from accurate experimental data, whereas 
near the threshold region the cross section
tends to be significantly small. Therefore,
such a study seemed to be irrelevant in the past. This
situation of course changes with the operation of 
precise hadron detectors in modern accelerators such as
CEBAF in Newport News and MAMI in Mainz.

In this paper I extend my previous analysis \cite{mart_thr,mart_k0lambda}
to the four isospin 
channels of  $K\Sigma$ photoproduction, i.e. the 
$K^+\Sigma^0$, $K^0\Sigma^+$, $K^+\Sigma^-$, and
$K^0\Sigma^0$ productions. This analysis becomes
part of the program for upgrading the phenomenological
kaon photo- and electroproduction model, i.e. the Kaon-Maid 
\cite{kaon-maid}. I also believe that it is important to study 
the processes in details, especially near the production thresholds,
where a number of unknown parameters can be easily fixed and 
many related but important aspects can be also studied. 
As an example, the electromagnetic form factor of kaon has been shown
to produce significant effects near the thresholds 
of the $K^+\Lambda$ and $K^0\Lambda$
channel \cite{mart_thr,mart_k0lambda}. 
Considering fewer uncertainties at this kinematics, my previous
analyses obviously encourage investigation of the kaon charge form factors 
at thresholds. 
Furthermore, by using the method developed in \cite{igor,arndt:2009}, 
small structure appearing at $W\approx 1.65$ GeV 
in the $K^+\Lambda$ polarization observable provides an important
evidence of a new missing resonance or a narrow resonance 
\cite{mart-narrow} predicted by the chiral quark soliton 
model \cite{diakonov}.

The four isospin channels of $K\Sigma$ photoproduction along with 
their threshold energies are given in Table \ref{tab:threshold}.
It is apparent that photoproduction of  $K\Sigma$ is similar
to the photoproduction of pion-nucleon ($\pi N$), 
because it involves the production
of isospin 1 and isospin 1/2 hadrons. However, 
the difference is also obvious, i.e. in the case of kaon (pion)
the $\Sigma$-hyperon ($\pi$-meson) has isospin 1, whereas the 
$K$-meson (nucleon) has isospin 1/2. 
Besides that, the presence of explicit strangeness in the 
case of kaon makes kaon photoproduction more unique than
pion photoproduction. 

There have been extensive discussions in the literature 
\cite{Kiswandhi:2011cq,Saghai:2006ra,Janssen:2001wk,
DeCruz:2012bv,Ireland:2004kp,Borasoy:2007ku,Klempt:2009pi,
Nikonov:2007br,anisovich_EPJA_2011,anisovich_EPJA_2012,
missing-d13,Maxwell:2012,DeCruz:2011xi}
about strangeness photoproduction which provides
a new mechanism to investigate the so-called missing resonances,
i.e. the resonances predicted by quark models but not listed
by the PDG \cite{olive}, since it does not appear 
in the pion-nucleon scattering process.
This is due to the fact that their decay widths are only sizable  
to the strangeness channels, rather than to the $\pi N$ 
channels \cite{capstick94}. An example of such efforts
has been performed for the $K^+\Lambda$ channel, where a
$D_{13}(1895)$ as a candidate of the missing nucleon
resonance \cite{missing-d13} was concluded from an analysis 
of the second peak in the cross section
of the $K^+\Lambda$ photoproduction data from
SAPHIR 1998 \cite{saphir98}. Note that different 
conclusion, however, could be drawn by using recent experimental 
data \cite{Janssen:2001wk}. Recently, it is found that
the peak originates mostly from the contribution of 
the  $P_{13}(1900)$, instead of the $D_{13}(1895)$
\cite{Nikonov:2007br,mart_p13}.

Furthermore, there is an intrinsic difference between photoproductions of 
$K\Sigma$ and $K\Lambda$, which comes from the consequence of
the hyperon isospin in the final states. Since $\Sigma$ 
is an isovector particle, photoproduction of 
$K\Sigma$ yields a total isospin 3/2 in the final state and, 
as a consequence, allows for isospin 3/2 ($\Delta$) intermediate 
states in the $s$-channel. Thus, 
photoproduction of $K\Sigma$ would provide more information
not available from photoproduction of $K\Lambda$. Although the number
of resonances increases with the inclusion of $\Delta$ resonances,
the total number of resonances used in the present investigation is only
four, in which only one $\Delta$ resonance is relevant, 
i.e. the $\Delta(1700)D_{33}$.

\begin{table}[!t]
  \centering
\caption{Threshold energies of the $K\Sigma$ photoproductions 
  off the proton in terms of the photon laboratory energy 
  ($E_\gamma^{\rm thr.}$) and the total c.m. energy ($W^{\rm thr.}$).
  }
  \label{tab:threshold}
  \begin{ruledtabular}
  \begin{tabular}[c]{clccc}
    No.&Channel & $E_\gamma^{\rm thr.}$ (MeV)&~~~~~~&$W^{\rm thr.}$ (MeV)\\
    \hline
    1& $ \gamma + p$ $\longrightarrow$  $K^{+} + \Sigma^{0}$ &  1046 && 1686\\
    2& $ \gamma + p$ $\longrightarrow$  $K^{0} + \Sigma^{+}$ &  1048 && 1687\\
    3& $ \gamma + n$ $\longrightarrow$  $K^{+} + \Sigma^{-}$ & 1052  && 1691\\
    4& $ \gamma + n$ $\longrightarrow$  $K^{0} + \Sigma^{0}$ & 1051  && 1690\\
  \end{tabular}
  \end{ruledtabular}
\end{table}

Photoproduction of $K\Sigma$ was first considered more than 50 years 
ago in the lowest order perturbation theory by exploiting 
very modest information on the spin, parity,  and coupling 
constants of both kaon and hyperon \cite{Kawaguchi}. 
By normalizing the leading coupling constants 
($g_{K\Lambda N}/\sqrt{4\pi}$ and $g_{K\Sigma N}/\sqrt{4\pi}$) 
to unity a number of cross sections
for different photon energies were calculated. At the same time, 
a similar calculation was also made with variation 
of coupling constants \cite{fujii}. 
However, investigation of $K\Sigma$ channels began
more attractive only after the finding of Ref.~\cite{Mart:1995wu},
which showed that the available phenomenological models 
for the $K^+\Sigma^0$ process over predict
the $K^0\Sigma^+$ cross section by almost two orders of magnitude. 
Only after including very few $K^0\Sigma^+$ data in the fit, this problem
can be partly alleviated. Unfortunately, the
extracted leading coupling constants are too small and cannot be reconciled 
with the prediction of SU(3) and those extracted from the kaon
scattering processes \cite{Mart:1995wu}. 
Since most of the contributions come from the
Born terms, introducing hadronic form factors in hadronic vertices of 
the scattering amplitude might become the suitable choice. 
There is a number of recipes
proposed in the literature for including these form factors without 
destroying gauge invariance of the process \cite{Haberzettl:1998eq}.
In spite of significant improvements in the model, the inclusion of
hadronic form factors simultaneously over-suppresses the cross section at 
very forward angles \cite{bydzovsky}. Furthermore, different
methods to suppress the excessively large Born terms 
also exist in the literature. For instance, 
Ref.~\cite{janssen-sigma} proposed the use of hyperon resonances, instead of
hadronic form factors, to
overcome the large contribution of background terms. Meanwhile,
within the framework of chiral quark model (CQM), Ref.~\cite{zhenping-sigma}
showed that the inclusion of higher-mass and spin resonances 
could also overcome 
the problem of the $K^0\Sigma^+$ cross section over prediction, since
the $\Delta(1905)F_{35}$, $\Delta(1910)P_{31}$,  $\Delta(1920)P_{33}$, 
and $\Delta(1950)F_{37}$ resonances have been shown to 
yield a minimum at $W\approx 1.9$ GeV 
\cite{zhenping-sigma}.

In contrast to the $K^+\Lambda$ channel, where the problem of 
data consistency has been severely plagued phenomenological 
analyses for years \cite{Mart:2006dk,Mart:2009nj}, 
experimental data of the $K^+\Sigma^0$
channel from SAPHIR 2004 \cite{Glander:2003jw}, CLAS 2006
\cite{Bradford:2005pt}, and CLAS 2010 \cite{biblap} seem to be 
consistent. Thus, the number of available experimental data near threshold
in this channel is relatively large, improving the accuracy of the
present analysis. 

The organization of this paper is as follows. In Sec. \ref{sec:formalism}
I briefly present the formalism used in my analysis. In Sec.
\ref{sec:formalism} I discuss the numerical result obtained for the
$K\Sigma$ photoproduction. Result for the electroproduction case
is given in Sec. \ref{result-electroproduction}. Section
\ref{sec:mami} is exclusively devoted to discuss the recent
MAMI electroproduction data at very low $Q^2$. 
Section \ref{sec:crystal-ball} presents the effect of the new
Crystal Ball data on my calculation. In Sec.
\ref{sec:form-factor} I briefly discuss the effect of the
$K^0$ charge form factor on the cross sections of the $K^0\Sigma^+$
and $K^0\Sigma^0$ channels. I will summarize my present analysis 
and conclude my finding in Sec. \ref{sec:summary}. A small portion
of the result obtained in the present analysis, that uses the older 
PDG information \cite{pdg2010}, has been presented in conferences 
\cite{mart:conferences,mart:conferences1}. 
In this paper I present the comprehensive
result of my analysis. Furthermore, here I use the information
of nucleon resonances obtained from the latest PDG report
\cite{olive}, which leads to a slightly different result
in the extracted nucleon resonance properties as well as the
calculated observables. 

\section{Formalism}
\label{sec:formalism}
A complete formalism of the background and resonance amplitudes 
for the $\gamma +p\to K^++\Lambda$ channel has been written 
in my previous paper \cite{mart_thr}. For use in the four channels
$K\Sigma$ photoproduction, a number of modifications is needed. 
This includes the the isospin relation of 
hadronic coupling constants in the background terms 
\cite{Mart:1995wu}, i.e.,
\begin{eqnarray}
\label{eq:su31}
g_{K^{+} \Sigma^{0} p} = -g_{K^{0} \Sigma^{0} n}
 = g_{K^{0} \Sigma^{+} p}/ \sqrt{2} = g_{K^{+} \Sigma^{-} n}/ \sqrt{2} ,\\
g_{K^{+} \Lambda p} = g_{K^{0} \Lambda n} ,~
g^{V,T}_{K^{*+} \Lambda p} = g^{V,T}_{K^{*0} \Lambda n} ,
\label{eq:su32}
\end{eqnarray}
as well as the isospin factor 
\begin{eqnarray}
  c_{K\Sigma} = \left\{ \begin{array}{ll}
    -1/\sqrt{3} & ; ~{\rm isospin}~ 1/2\\
    \sqrt{3/2} & ; ~{\rm isospin}~ 3/2
    \end{array}\right. 
\end{eqnarray}
of the multipoles  \cite{hanstein99} in the resonance terms, i.e.,
\begin{equation}
  A_{\ell\pm}^R(W) = {\bar A}_{\ell\pm}^R
  \, c_{K\Sigma }\, \frac{f_{\gamma R}(W)\, 
    \Gamma_{\rm tot}(W) M_R\, f_{K R}(W)}{M_R^2-W^2-iM_R\Gamma_{\rm tot}(W)}~ e^{i\phi} ,
  \label{eq:m_multipole}
\end{equation}
where the total width $\Gamma_{\rm tot}$  can be related to
the resonance width ($\Gamma_R$) by using Eq.~(11) of Ref.~\cite{Mart:2006dk}.
More detailed explanation of Eq.~(\ref{eq:m_multipole}) can be found in
Section II of Ref.~\cite{Mart:2006dk}.

In the case of resonance contribution I adopt the convention of pion 
photo- and electroproduction \cite{hanstein99} for
the physical amplitudes of the kaon photo- and electroproduction, 
\begin{eqnarray}
  \label{eq:physical-amplitudes1}
  A(\gamma+ p\to K^++\Sigma^0) &=& A_{p}^{(1/2)}+{\textstyle\frac{2}{3}}\, A^{(3/2)},\\
  \label{eq:physical-amplitudes2}
  A(\gamma+ p\to K^0+\Sigma^+) &=& \sqrt{2}\left[
    A_{p}^{(1/2)}-{\textstyle\frac{1}{3}}\, A^{(3/2)}\right],\\
  \label{eq:physical-amplitudes3}
  A(\gamma+ n\to K^++\Sigma^-) &=& \sqrt{2}\left[
    A_{n}^{(1/2)}+{\textstyle\frac{1}{3}}\, A^{(3/2)}\right],\\
  A(\gamma+ n\to K^0+\Sigma^0) &=& -A_{n}^{(1/2)}+{\textstyle\frac{2}{3}}\, A^{(3/2)},
  \label{eq:physical-amplitudes4}
\end{eqnarray}
where $A_{p}^{(1/2)}$ and $A_{n}^{(1/2)}$ are the proton and neutron amplitudes
with total isospin 1/2, respectively, whereas $A^{(3/2)}$ is the amplitude for
the isospin 3/2 contribution. The formalism given by 
Eqs.~(\ref{eq:physical-amplitudes1})--(\ref{eq:physical-amplitudes4})
can be implemented in the calculation of CGLN amplitudes $F_1,\cdots,F_6$ 
\cite{hanstein99} from the multipoles before 
calculating the cross section or polarization observables.
Note that by limiting the energy up to 50 MeV above the production
thresholds the number of electric, magnetic, and scalar multipoles
are significantly limited. As a consequence, the relation between
CGLN amplitudes and the multipoles can be simplified to
\begin{eqnarray}
  \label{eq:f1}
  F_1 &=& E_{2-}+3M_{2-}+\, 3\left( E_{1+}+M_{1+}\right)\, \cos\theta  ,\\
  \label{eq:f2}
  F_2 &=& 2M_{1+}+M_{1-}+6M_{2-} \,\cos\theta  ,\\
  \label{eq:f3}
  F_3 &=& 3\left(\, E_{1+}-M_{1+}\right)  ,\\
  \label{eq:f4}
  F_4 &=& -3\left(\, M_{2-}+E_{2-}\right) ,\\
  \label{eq:f5}
  F_5 &=& S_{1-}-2S_{1+}+6\, S_{2-}\,\cos\theta ,\\
  \label{eq:f6}
  F_6 &=& 6\, S_{1+}\,\cos\theta-2\, S_{2-}~.
\end{eqnarray}

Since both proton and neutron channels exist in $K\Sigma$
photoproduction, I obviously need the ratios between
charged and neutral kaon transition moments 
$r_{K^*K\gamma}\equiv g_{K^{*0} K^{0} \gamma}/g_{K^{*+} K^{+} \gamma}$
and $r_{K_1K\gamma}\equiv g_{K_1^{0} K^{0} \gamma}/g_{K_1^{+} K^{+} \gamma}$.
The first ratio can be fixed by using the PDG values \cite{olive},
whereas the second ratio can be considered as a free parameter
during the fit process, because, fortunately, one of the proton channels 
that produces neutral kaon ($\gamma+p\to K^0+\Sigma^+$) 
has experimental data, though very limited. More detailed discussion
about this topic will be given when I discuss the result of the
$K^0\Sigma^+$ channel in Sect.~\ref{sec:result}.

\begin{figure}[t]
  \begin{center}
    \leavevmode
    \epsfig{figure=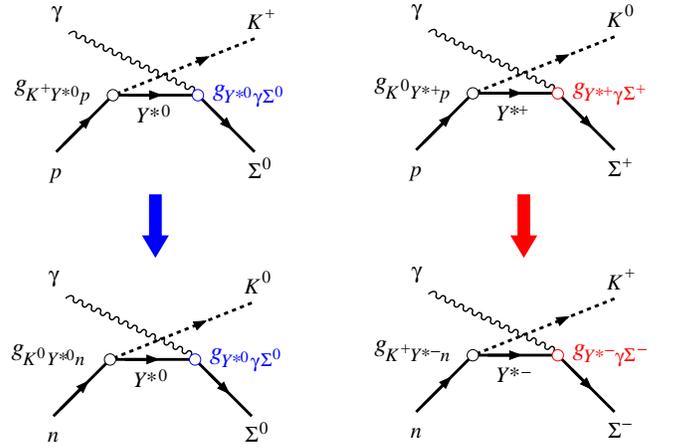,width=85mm}
    \caption{(Color online) Hadronic and electromagnetic
      coupling constants in the hyperon resonance intermediate state
      of $K\Sigma$ photoproductions. Whereas the hadronic coupling
      constants are related through Eq.~(\ref{eq:su31}), the electromagnetic
      coupling constants of the charged $\Sigma$ production is different
      from that of the neutral $\Sigma$ production.}
   \label{fig:ystar} 
  \end{center}
\end{figure}

Note that the electromagnetic vertices of 
hyperon resonances in the charged hyperon productions
($K^0\Sigma^+$ and $K^+\Sigma^-$) are different from 
those in the neutral hyperon productions 
($K^+\Sigma^0$ and $K^0\Sigma^0$). Figure~\ref{fig:ystar}
exhibits the corresponding coupling constants for all
$\Sigma$ channels. The hadronic part of coupling constants $g_{KY^*N}$ are 
obviously related by using Eq.~(\ref{eq:su31}), 
whereas the electromagnetic coupling $g_{Y^*\gamma\Sigma}$
depends on the charge of the hyperon. Obviously, the neutral hyperon
($\Sigma^0$) productions use the same $g_{Y^{*0}\gamma\Sigma^0}$ coupling,
while the charged hyperon ($\Sigma^+, \Sigma^-$) productions 
use the same $g_{Y^{*+}\gamma\Sigma^+}$ coupling. Thus, in the fitting
process one can use the ratio 
$c_{Y^*}\equiv g_{Y^{*+}\gamma\Sigma^+}/g_{Y^{*0}\gamma\Sigma^0}$
as a free parameter in order to distinguish the charged
hyperon resonance from the neutral one.

Note that the above formalism is also valid in the case of
electroproduction. To this end I use the standard 
electromagnetic form factors as in my previous work 
\cite{mart_thr,mart_k0lambda} for extension of my model
to the finite $Q^2$ region, where $Q^2$ is the virtual
photon momentum squared. As in the previous work \cite{mart_thr} I do not
use the hadronic form factor in the present work, because the considered
energy is sufficiently low and, as a consequence, the agreement between 
model calculation and experimental data can be easily achieved.

\begin{table*}[tb]
  \centering
  \caption{Properties of the nucleon resonances used in the present
    analysis \cite{olive}. $M_R$ and $\Gamma_R$ are the mass and width
  of the resonance, respectively, $A_{1/2}$ and $A_{3/2}$ are the
  resonance photo-decay helicity amplitudes, 
  and $\beta_{K\Lambda}$ is the 
  kaon branching ratio to the $K\Lambda$ channel. See Sect. III of
  Ref.~\cite{mart_thr} for further explanation of these  
  Breit-Wigner resonance parameters. The status of the resonance is given by a number of asterisks (*) according to the PDG. Further explanation of this status can be found in Ref.~\cite{olive}.}
  \label{tab:resonance_pdg}
  \begin{ruledtabular}
  \begin{tabular}[c]{lcccc}
    Resonance&$N(1700)D_{13}$&$\Delta(1700)D_{33}$&$N(1710)P_{11}$&$N(1720)P_{13}$\\
    \hline
    $M_R$ (MeV)&$1700\pm 50$&$1700^{+50}_{-30}$ &$1710\pm 30$&$1720^{+30}_{-20}$ \\
    $\Gamma_R$ (MeV)&$150^{+100}_{-50}$&$300\pm 100$&$100^{+150}_{-50}$ &$250^{+150}_{-100}$\\
    $\beta_{K\Lambda}$&$<0.03$&$\cdots$&$0.15\pm 0.10$   &$0.044\pm 0.004$ \\
    $A_{1/2} (p)$ ($10^{-3}$GeV$^{-1/2}$)&$15\pm 25$ &$+140\pm 30$&$+40\pm 20$&$+100\pm 20$\\
    $A_{3/2} (p)$ ($10^{-3}$GeV$^{-1/2}$)& $-15\pm 25$ &$+140\pm 30$ &$\cdots$&$150\pm 30$\\
    $A_{1/2} (n)$ ($10^{-3}$GeV$^{-1/2}$)&$20\pm 15$    &$+140\pm 30$ & $-40\pm 20$ & $+7\pm 15$\\
    $A_{3/2} (n)$ ($10^{-3}$GeV$^{-1/2}$)&$-30\pm 20$ &$+140\pm 30$  & $\cdots$ & $-5\pm 25$\\
    Overall status&*** &****&*** &****\\
    Status seen in $K\Sigma$ & * & * & * & * \\
  \end{tabular}
  \end{ruledtabular}
\end{table*}

As in the previous study I consider the energies from the production
threshold ($W\simeq 1690$ MeV) up to 50 MeV above the threshold
($W\simeq 1740$ MeV). In this energy range there are three nucleon and
one $\Delta$ 
resonances in the PDG listing, i.e. the $N(1700)D_{13}$, $\Delta(1700)D_{33}$, 
$N(1710)P_{11}$, and $N(1720)P_{13}$ resonances. Their properties
relevant to the present study are listed in Table \ref{tab:resonance_pdg}.
Properties of the particles as well as other parameters used 
in the background terms can be found
in my previous work \cite{mart_thr}. Due to the nature of resonance
formalism used in the present study \cite{hanstein99}, the nucleon 
resonances with masses below the thresholds energy cannot be
included as in
the case of covariant formalism \cite{Mart:coupl_strength}.

There are 331 experimental data points in the database, dominated by the
$K^+\Sigma^0$ photoproduction differential cross section 
\cite{biblap,Glander:2003jw,Bradford:2005pt,McNabb2004}.
In addition, there are also data for the $K^+\Sigma^0$ electroproduction 
differential cross section \cite{Ambrozewicz:2006zj}, 
$K^0\Sigma^+$ photoproduction differential cross section 
\cite{lawall}, $K^+\Sigma^0$ recoil polarization 
\cite{biblap,Glander:2003jw}. Other data, such as from the
$K^+\Sigma^0$ and 
$K^+\Sigma^-$ photoproduction measurement by LEPS collaboration
\cite{sumihama06,kohri06}, have photon energies beyond 
the upper limit.
Note that the number of data points in the $K\Sigma$ production
near threshold is larger than that in the $K\Lambda$ case 
\cite{mart_thr} (139 data points). Thus, a better 
statistics obtained from the present analysis could be expected.

\section{Results and Discussion}
\label{sec:result}

The extracted parameters from fit to 331 data points are 
displayed in Tables \ref{tab:background} and \ref{tab:resonance}, 
where the background and resonance parameters are  separated
in different Tables for the sake of convenience. Note that 
the effect of the $\Lambda(1800)S_{01}$ 
resonance, which was found to play important role in the $K^+\Lambda$ 
photoproduction near threshold \cite{mart_thr}
as well as at higher photon energies \cite{Mart:2013hia},
have also been investigated. 
In the present work it is found that the effect is almost
negligible, i.e. the $\chi^2/N$ is reduced from 1.07 
to 1.05, whereas other extracted parameters do not dramatically change 
after including the $\Lambda(1800)S_{01}$ resonance 
(see Table \ref{tab:background}). Therefore, in the following 
discussion the hyperon resonance will be excluded. 

Table \ref{tab:background} indicates that contributions of the
$K^*(892)$ and $K_1(1270)$ vector mesons are relatively small,
which is in contrast to the case of $K\Lambda$ production \cite{mart_thr}.
The extracted ratio $r_{K_1K\gamma} = 2.99$ is obviously larger than that 
obtained in Kaon-Maid, i.e., $-0.45$. The extracted $r_{K_1K\gamma}$ 
decreases when the  $\Lambda(1800)S_{01}$ hyperon resonance 
is included in the model (see Table \ref{tab:background}).
This issue will be discussed later in this Section. 

The achieved 
$\chi^2$ per number of degrees of freedom is close to one,
indicating that the omission of hadronic form factors in the present
analysis does not lead to a serious problem as in the analyses
beyond the threshold region. The extracted resonance properties
shown in Table \ref{tab:resonance} do not show any dramatic
deviations from the PDG values, since during the fit 
they were varied within
the error bars given by the PDG \cite{olive}. 

\begin{table}[!h]
  \centering
  \caption{Extracted background parameters from fit to experimental data
    by excluding and including the $\Lambda(1800)S_{01}$ 
    hyperon resonance (indicated by $Y^*$).
    Note that during the fits the main coupling constants, 
    $g_{K \Lambda N}/\sqrt{4\pi}$ and $g_{K \Sigma N}/\sqrt{4\pi}$,
    were varied within the values accepted by the SU(3) prediction
    with a 20\% symmetry breaking \cite{Adelseck:1990ch}.
    }
  \label{tab:background}
  \begin{ruledtabular}
  \begin{tabular}[c]{lrr}
    Coupling Constants & Without $Y^*$ & With $Y^*$  \\
\hline
  $g_{K \Lambda N}/\sqrt{4\pi}$&$-3.18$   & $-3.00$ \\
  $g_{K \Sigma N}/\sqrt{4\pi}$ &$ 1.30$   & 1.30 \\
  $G^{V}_{K^{*}}/4\pi$         &$-0.02$   & $-0.04$\\
  $G^{T}_{K^{*}}/4\pi$         &$-0.32$   & $-0.32$\\
  $G^{V}_{K_1}/4\pi$           &$-0.03$   & $-0.14$ \\
  $G^{T}_{K_1}/4\pi$           &$-0.04$   & $-0.32$ \\
  $G_{Y^*}/4\pi$               & $\cdots$ & $-1.70$\\
  $r_{K_1K\gamma}$             & $2.99$  & $2.07$\\
  $c_{Y^*}$                    & $\cdots$ & $1.33$\\
  $\Lambda_{K}$ (GeV)          & $ 1.63$  & 1.63\\
  $\Lambda_{K^*}$ (GeV)        & $ 0.50$  & 0.50\\
  $\Lambda_{K_1}$ (GeV)        & $ 0.50$  & 0.50\\
  $\Lambda_{Y}$ (GeV)          & $ 0.50$  & 0.50\\
  \hline
  $\chi^2/N$ &1.07 & 1.05 \\
  \end{tabular}
  \end{ruledtabular}
\end{table}

\begin{table*}[!]
  \centering
  \caption{Extracted resonance parameters from fit to experimental data.
  $\beta_{K\Sigma}$ is the kaon branching ratio to the $K\Sigma$ channel,
  $\phi$ is a Breit-Wigner resonance parameter given in Eq.~(7) of 
  Ref.~\cite{mart_thr}, whereas $\alpha_{N^*}$ and $\beta_{N^*}$
  are the parameters of the $Q^2$ dependence of the 
  resonance multipoles given by Eq.~(\ref{eq:Q2_dependence})
  in Sect.~\ref{result-electroproduction}.}
  \label{tab:resonance}
  \begin{ruledtabular}
  \begin{tabular}[c]{lcccc}
    Resonance&$N(1700)D_{13}$&$\Delta(1700)D_{33}$&$N(1710)P_{11}$&$N(1720)P_{13}$\\
\hline
  $M_R$ (MeV)                           & 1716 & 1692 &  1727  & 1700 \\
  $\Gamma_R$ (MeV)                      & 250  & 400  &   50   &  150  \\
  $\beta_{K\Sigma}$                     & $3.0\times 10^{-2}$&$6.4\times 10^{-5}$&  $5.5\times 10^{-3}$ & $1.2\times 10^{-4}$\\ 
  $\phi$ (deg)                          & $163$& $63$ &  $40$  &$360$\\
  $A_{1/2}(p)$ ($10^{-3}$ GeV$^{-1/2}$) & $40$ & $170$&  43    & 120   \\
  $A_{3/2}(p)$ ($10^{-3}$ GeV$^{-1/2}$) & $-6$ & $110$ &$\cdots$ & 120  \\
  $S_{1/2}(p)$ ($10^{-3}$ GeV$^{-1/2}$) & $-27$& $-100$&  $26$ &$-100$    \\
  $\alpha_{N^*}$ (GeV$^{-2}$)           &0.00  &3.86  &2.98    &9.93\\
  $\beta_{N^*}$ (GeV$^{-2}$)            &1.44  &3.07  &1.98    &2.77\\
  \end{tabular}
  \end{ruledtabular}
\end{table*}

\begin{figure}[!t]
  \begin{center}
    \leavevmode
    \epsfig{figure=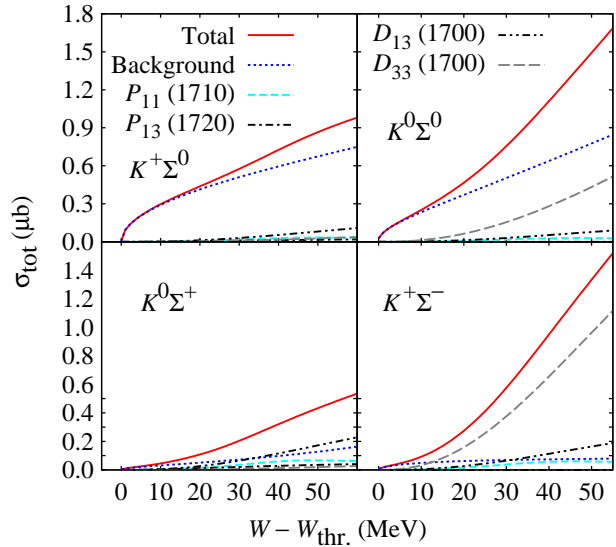,width=85mm}
    \caption{(Color online) Contribution of the background,  
      $N(1710)P_{11}$, $N(1720)P_{13}$, $N(1700)D_{13}$, and $\Delta(1700)D_{33}$ 
      resonance amplitudes to the total cross section
      of the $\gamma+N\to K\Sigma$ processes in four isospin channels. }
   \label{fig:contrib} 
  \end{center}
\end{figure}

Comparison between contributions of the background and
resonance terms is displayed in Fig.~\ref{fig:contrib}.
It is obvious from this figure that contribution of the background
terms is dominant in the $K^+\Sigma^0$ and $K^0\Sigma^0$ channels,
which can be understood as the isospin effects in the
background amplitudes given by Eqs.~(\ref{eq:su31}) and 
(\ref{eq:su32}). In contrast to this, the effect of resonances 
clearly shows up in both $K^0\Sigma^+$ and $K^+\Sigma^-$
channels. This phenomenon is also understood from the isospin factors
in the resonances given by Eqs.~(\ref{eq:physical-amplitudes2}) and 
(\ref{eq:physical-amplitudes3}). 

\begin{figure}[t]
  \begin{center}
    \leavevmode
    \epsfig{figure=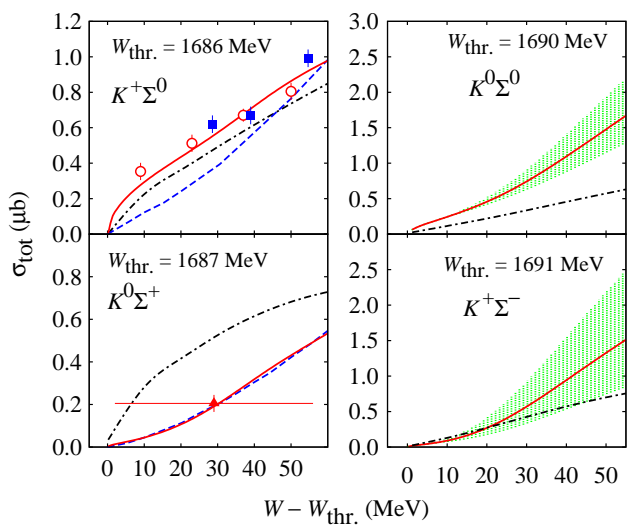,width=85mm}
    \caption{(Color online) Total cross section
      obtained in the present work (solid lines) compared with 
      the results of Kaon-Maid \cite{kaon-maid} (dashed-dotted lines)
      and chiral perturbation theory (CHPT) \cite{Steininger:1996xw} 
      (dashed lines) as well as the available experimental data 
      from the SAPHIR collaboration (open circles \cite{Glander:2003jw} 
      and solid triangle \cite{lawall}), and  the CLAS 
      collaboration (solid squares \cite{Bradford:2005pt}). In the case of
      $K^0\Sigma^0$ and $K^+\Sigma^-$ channels (right panels)
      the uncertainties
      of the present calculations, due to the uncertainties
      in the helicity photon couplings of the 
      resonances as given in Table~\ref{tab:resonance_pdg},
      are indicated by the shaded green areas. If these 
      uncertainties are excluded, the result is shown by the
      solid lines. Note that both the present work and Kaon-Maid
      do not include the total cross section data shown 
      in this figure in the fitting database. The CHPT uses
      the leading coupling constants predicted by SU(3) 
      as the input for calculating this cross section.}
   \label{fig:total4}
  \end{center}
\end{figure}

Comparison between the predicted total cross sections and those from previous
works \cite{Steininger:1996xw,kaon-maid} as well as experimental 
data \cite{Glander:2003jw,Bradford:2005pt,lawall} is shown in Fig.~\ref{fig:total4}. 
It is obvious that the present analysis provides a more accurate
prediction in both proton channels ($K^+\Sigma^0$ and $K^0\Sigma^+$).
In the neutron channels ($K^+\Sigma^-$ and $K^0\Sigma^0$) the 
cross section uncertainties are found to be relatively large
as shown by the shaded area in the right panels of  
Fig.~\ref{fig:total4}, especially at $W\approx 1700$ MeV, 
where most of the involved resonances  are located. Note
that all uncertainties shown in the right panels of Fig.~\ref{fig:total4}
originates from the uncertainties in the neutron helicity amplitudes 
$A_{1/2}(n)$ and $A_{3/2}(n)$ given by PDG (see Table 
\ref{tab:resonance_pdg}) \cite{olive}. 
Thus, experimental data in both neutron channels
are urgently required to reduce this uncertainty. Such experimental data
could be expected from the $K^0$ photoproduction experiment performed
by the Tohoku group which uses deuteron as a target \cite{tsukada}.
Nevertheless, for this purpose, higher statistics data are more recommended 
in order to reduce
some uncertainties coming from Fermi motion in the deuteron 
as well as from the final-state interaction induced by the spectator 
nucleon. For the $K^+\Sigma^-$ photoproduction off a deuteron 
experimental data have been available from the CLAS collaboration
with photon lab energies from 1.1 GeV (almost 50 MeV above the threshold,
see Table \ref{tab:threshold}) up to 3.6 GeV \cite{pereira}. 
Although the lowest energy is very close to the upper limit of the
present analysis,
the challenging task now is to remove the effects of initial- 
and final-state interactions from the data.

\begin{figure}[!h]
  \begin{center}
    \leavevmode
    \epsfig{figure=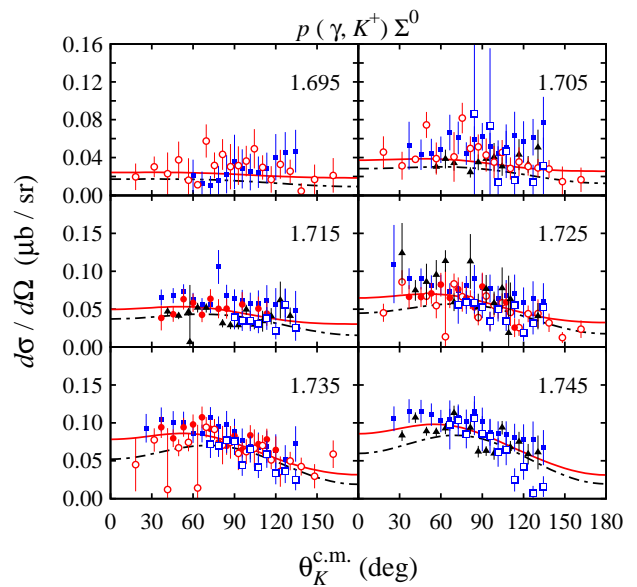,width=85mm}
    \caption{(Color online) Comparison between angular distributions of 
             the $\gamma+p\to {K^+}+\Sigma^0$
             differential cross section 
             obtained from the present and Kaon-Maid 
             \cite{kaon-maid} models with experimental data 
             from the SAPHIR (open circles \cite{Glander:2003jw}), 
             CLAS (solid squares \cite{biblap}, solid circles
             \cite{Bradford:2005pt}, and solid triangles \cite{McNabb2004}),
             and Crystal Ball (open squares \cite{Jude:2013jzs}) 
             collaborations.
             The corresponding total c.m. energy $W$ (in GeV) is
             shown in each panel. Except the new Crystal Ball data, 
             all experimental data displayed 
             in this figure were used
             in the fit to obtain the solid line.}
   \label{fig:dif_th} 
  \end{center}
\end{figure}

Comparison between the calculated differential cross section of
the $K^+\Sigma^0$ channel with the prediction of Kaon-Maid \cite{kaon-maid} 
and experimental data \cite{Glander:2003jw,Bradford:2005pt,biblap} is
shown in Fig. \ref{fig:dif_th}. Within the existing experimental error bars
the present work also provides a significant improvement to
the result of Kaon-Maid, especially at $W=1735$ and 1745 MeV. Further
improvement can be also observed in the forward directions.
Note that in Kaon-Maid the problem in this kinematics originates 
from the inclusion of hadronic form factors, that over suppresses
the $K\Lambda$ cross sections at forward angles \cite{bydzovsky}.
Therefore, the present study also emphasizes the need for a thorough
investigation of the effects of including hadronic form factors on
differential cross sections at forward kinematics.

\begin{figure}[!h]
  \begin{center}
    \leavevmode
    \epsfig{figure=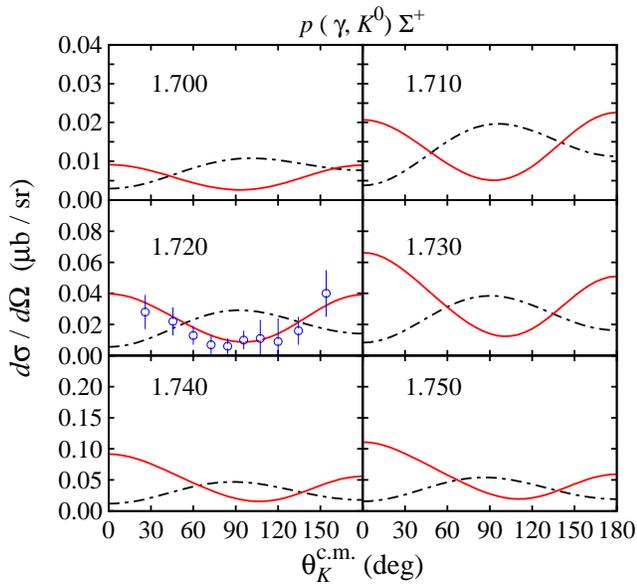,width=85mm}
    \caption{(Color online) Angular distribution of 
      the $\gamma+p\to {K^0}+\Sigma^+$ differential cross
      sections. Notation of the curves is as in 
      Fig.~\ref{fig:dif_th}. Experimental data
      at $W=1720$ MeV are from Ref.~\cite{lawall}.}
   \label{fig:dif_k0sp} 
  \end{center}
\end{figure}

\begin{figure}[!h]
  \begin{center}
    \leavevmode
    \epsfig{figure=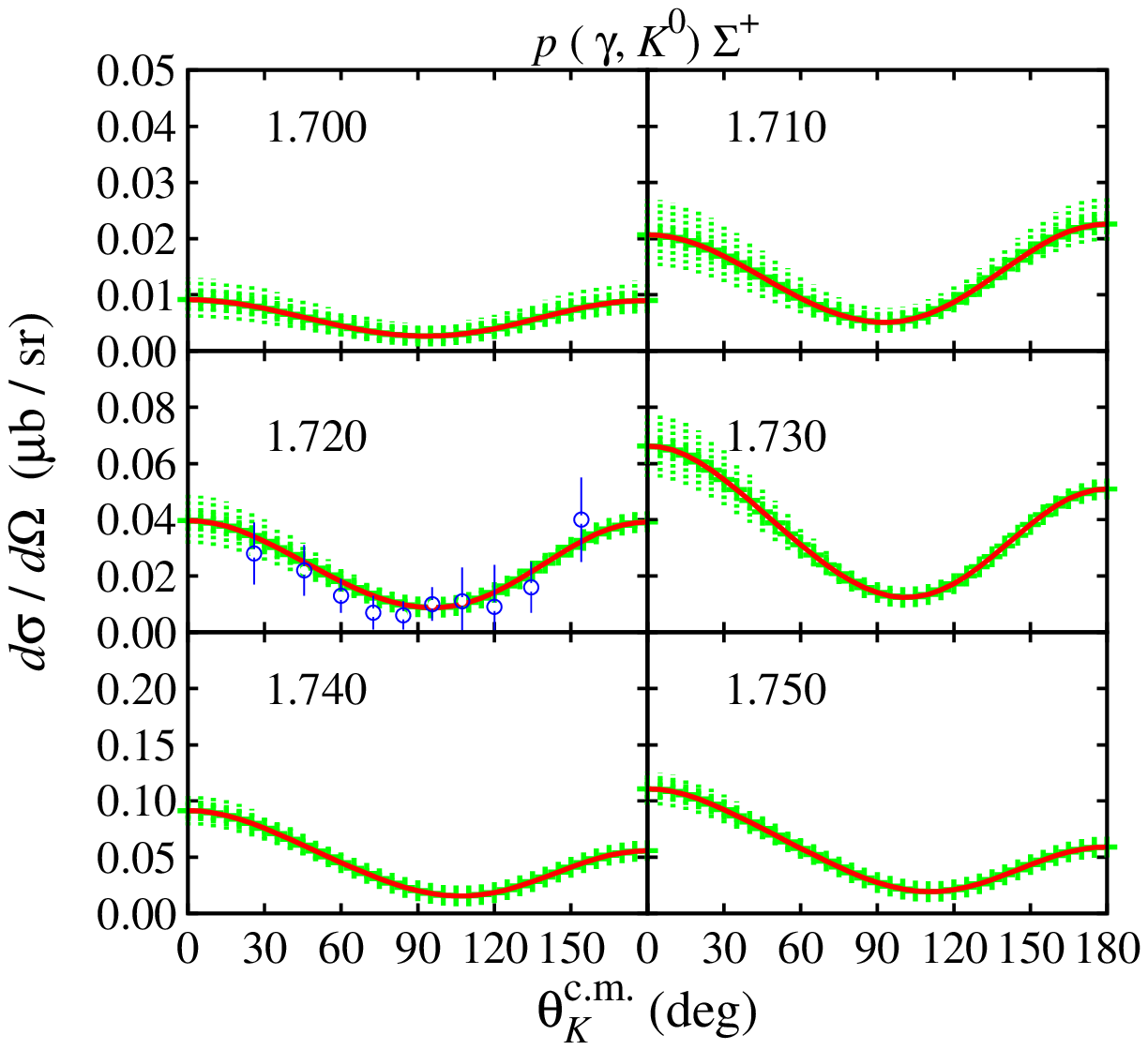,width=85mm}
    \caption{(Color online) Variation in the differential cross section
      of the $\gamma+p\to {K^0}+\Sigma^+$ process as a result of 
      20\% variation of the $r_{K_1K\gamma}$ values. Solid lines indicate
      the cross section calculated without this variation. Experimental data
      are as in Fig.~\ref{fig:dif_k0sp}.}
   \label{fig:dif_k0sp_var} 
  \end{center}
\end{figure}

Differential cross sections of the $\gamma+p\to {K^0}+\Sigma^+$ 
process display an interesting result. Unlike the prediction
of Kaon-Maid, which is almost similar to the $K^+\Lambda$ case,
here the cross sections rise sharply in the backward directions and
reach the minima at $\theta_K\simeq 90^\circ$. The cross section enhancement 
is also detected at forward angles. The result indicates a 
strong $u$-channel contribution which is solely mediated by 
the $\Sigma^+$ (since the $\Sigma^*$ resonances do not significantly
contribute and therefore are not included in the model) 
as well as important contributions from the $t$-channel 
intermediate states obtained from the $K^{0*}$ and $K_1^0$
meson resonances.

Photoproduction of ${K^0}\Sigma^+$ is especially important 
for the extraction of the ratio between the 
$K_1(1270)$ transition moments in $K^0$ and $K^{+}$ productions, i.e.
\begin{eqnarray}
r_{K_1K\gamma}\equiv
g_{K_1^{0} K^{0} \gamma}/g_{K_1^{+} K^{+} \gamma},
\label{eq:ratio_K1}
\end{eqnarray}
since there is no information available for the 
$K_1 \rightarrow K \gamma$ decay width. 
This is in contrast to the lower mass vector meson, the $K^*(892)$, 
since PDG provides the values of both $K^{+*} \rightarrow K^+ \gamma$
and $K^{0*} \rightarrow K^0 \gamma$ decay widths \cite{olive},
which can be related to their transition strengths 
by means of \cite{thom}
\begin{eqnarray}
\Gamma_{K^{*} \rightarrow K \gamma} = \frac{\rm 9.8~MeV}{4 \pi} \vert
g_{K^{*} K \gamma} \vert^{2} ,
\label{ksdecay}
\end{eqnarray}
whereas its sign can be constrained by using the cloudy bag model 
computation by Singer and Miller \cite{sing}.

In the present work the value of $r_{K_1K\gamma}$ 
is found to be $2.99$ 
(see Table~\ref{tab:background}). In Kaon-Maid this value was
obtained to be $-0.45$ \cite{mart_k0lambda}. 
The presently extracted  $r_{K_1K\gamma}$ is larger probably
because the contribution of $K_1(1270)$ would be different
for different models. The discrepancy between the two values
could originate from the small number of $K^0\Sigma^+$ 
data used in the database. 
In the present work the available data for the $\gamma+ p \to K^0+\Sigma^+$ 
channel near threshold are merely 10 points as shown in 
Fig.~\ref{fig:dif_k0sp}, whereas Kaon-Maid 
used only 29 data points in its database. 
As shown in Ref.~\cite{mart_k0lambda} this ratio is required
for extending the $K^+\Lambda$ photoproduction model to include the
$K^0\Lambda$ channel. In the pseudoscalar theory it was found
that variation of this ratio changes the cross section only 
at higher energies. However, the effect
is substantially large in the case of pseudovector coupling. 
It is also apparent that by
including the new Crystal Ball data
(will be discussed in Sect.~\ref{sec:crystal-ball}) 
a smaller ratio, i.e., $r_{K_1K\gamma}=2.07$, would be obtained,
which is in principle approaching the value of Kaon-Maid. 
This happens presumably
because the new Crystal Ball data are closer to the SAPHIR data,
instead of the CLAS ones (See Sect.~\ref{sec:crystal-ball}),
whereas Kaon-Maid was fitted to the the SAPHIR data \cite{saphir98}.

In order to investigate the sensitivity of the calculated cross sections 
depicted in Fig.~\ref{fig:dif_k0sp} to the $r_{K_1K\gamma}$ ratio, the
calculated cross sections are replotted in  Fig.~\ref{fig:dif_k0sp_var}.
Figure~\ref{fig:dif_k0sp_var} shows that 
the cross section changes if the ratio is varied 
within $\pm 20\%$. Obviously, sizable variations can be observed
at the forward and backward angles. Nevertheless, the presently
available data cannot resolve this variation. Therefore, it is urgent to 
measure this channel in order to improve our understanding on
the $K\Sigma$ photoproduction. With about $10\%$ error bars the 
experimental data would be able to constrain this ratio to vary 
within less than $20\%$ of its value.
Meanwhile, photoproduction experiment has been performed off a proton 
target by the CLAS collaboration at JLab. 
Data with very high statistics have been collected
and will be analyzed in the near future \cite{schumacher1}. 
Precise data on the $\gamma +p \to K^0+\Sigma^+$ channel would
allow us to extract not only the $r_{K_1K\gamma}$ ratio, but
also the corresponding ratio for the $K^*(892)$ vector meson.
Therefore, a more stringent constraint could be also applied to 
both $K^{+*} \rightarrow K^+ \gamma$ 
and $K^{0*} \rightarrow K^0 \gamma$ decay widths.

\begin{figure}[!h]
  \begin{center}
    \leavevmode
    \epsfig{figure=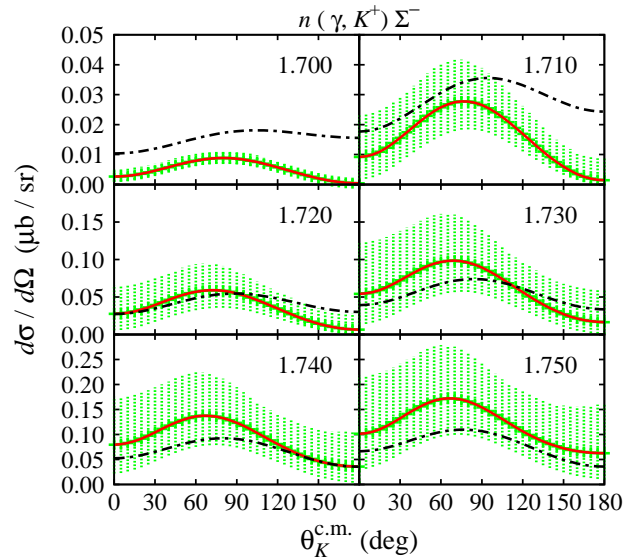,width=85mm}
    \caption{(Color online) As in Fig.~\ref{fig:dif_k0sp} but for
      the $\gamma+n\to {K^+}+\Sigma^-$ channel. The shaded areas 
      display the uncertainties
      of the present calculations due to the uncertainties
      in the helicity photon couplings as in Fig.~\ref{fig:total4}.}
   \label{fig:dif_kpsm} 
  \end{center}
\end{figure}

\begin{figure}[!h]
  \begin{center}
    \leavevmode
    \epsfig{figure=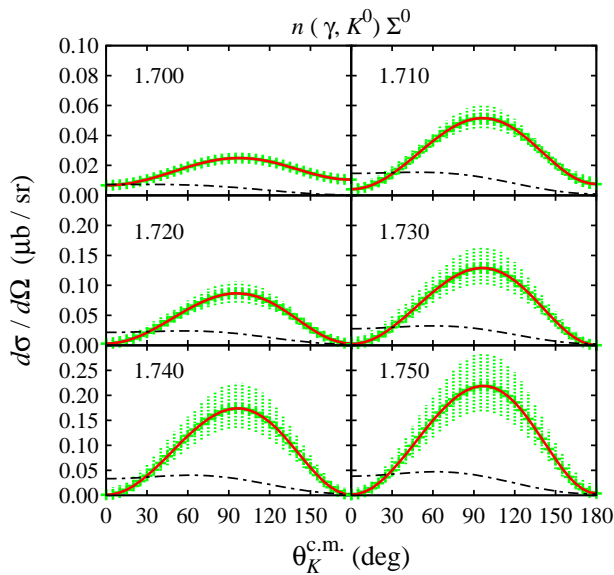,width=85mm}
    \caption{(Color online) As in Fig.~\ref{fig:dif_kpsm} but for
      the $\gamma+n\to {K^0}+\Sigma^0$ channel.}
   \label{fig:dif_k0s0} 
  \end{center}
\end{figure}

Figures \ref{fig:dif_kpsm} and \ref{fig:dif_k0s0} display the predicted
differential cross section of the $K^+\Sigma^-$ and $K^0\Sigma^0$
channels, respectively, where the prediction of Kaon-Maid
is also shown for comparison. In the
$K^+\Sigma^-$ channel, except for the lowest energy, the predictions of both 
Kaon-Maid and the present work are in agreement with each other,
whereas in the case of the $K^0\Sigma^0$ both predictions look very
different in the whole energy range. In both cases, however, the
present work indicates smaller uncertainties can be obtained
at energies very close to the threshold. Therefore, 
in both channels measurement
of the cross section close to the threshold is strongly recommended
to further constrain the present model. The energy dependent 
uncertainty exhibited
by the total cross section given in Fig.~\ref{fig:total4} can be
understood from the uncertainties shown in 
Fig. \ref{fig:dif_kpsm} and \ref{fig:dif_k0s0}.

The $\Sigma^0$ hyperon decays to a photon and a $\Lambda$ hyperon.
By analyzing the corresponding magnetic dipole ($M1$) 
transition matrix element,
which is proportional to $\bvec{\sigma}\cdot\bvec{\epsilon}$,
where $\bvec{\sigma}$ and $\bvec{\epsilon}$ are the Pauli matrix 
and photon polarization vector, respectively, it can be shown that 
the polarization of $\Lambda$ ($P_\Lambda$) is related to the
polarization of $\Sigma^0$ ($P_\Sigma$) through \cite{schumacher}
\begin{equation}
  \label{eq:polar_relation}
  P_\Lambda = -\frac{1}{3} \, P_\Sigma . 
\end{equation}
Experimental data on $K^+\Lambda$ and $K^+\Sigma^0$
photoproduction seem to obey this relation. 

\begin{figure}[!h]
  \begin{center}
    \leavevmode
    \epsfig{figure=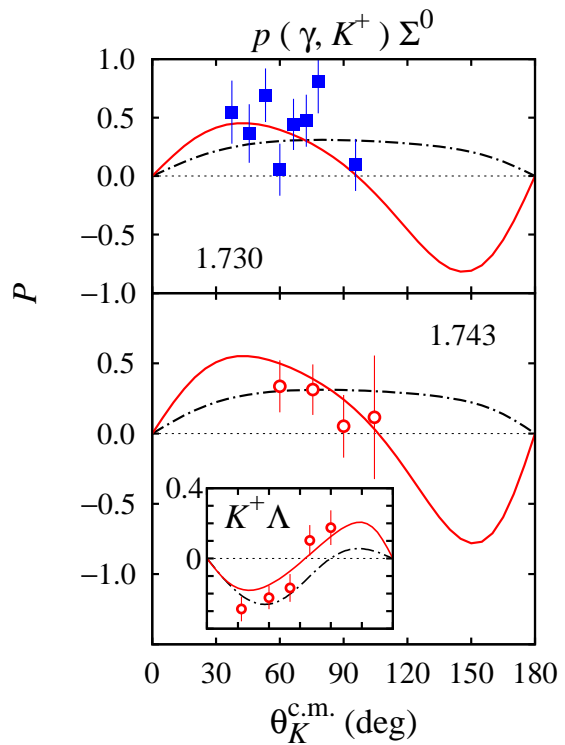,width=75mm}
    \caption{(Color online) Recoil polarization in the 
      $\gamma+p\to {K^+}+\Sigma^0$ process as a function of
      kaon scattering angle.
      Solid squares display the new CLAS data \cite{biblap}, whereas
      open circles exhibit the SAPHIR data \cite{Glander:2003jw}.
      Notation of the curves is as in Fig.~\ref{fig:dif_th}. 
      The inserted panel shows comparison between the previous
      calculation with Kaon-Maid and experimental data for
      the $\gamma + p \to K^+ +\vec{\Lambda}$ process \cite{mart_thr}.}
   \label{fig:dif_polar} 
  \end{center}
\end{figure}

As shown in the previous paper \cite{mart_thr}, $P_\Lambda$ 
exhibits an inverted sine function, i.e. it 
has negative values near the forward angle, but changes the 
sign near the  backward angle. 
Therefore, in the case of $\Sigma^0$ the expected polarization 
should display a sine function which is shown in Fig.~\ref{fig:dif_polar},
where it is obvious that the relation nicely works.
The agreement of experimental data with the present work is
probably not so surprising, because the data shown in 
Fig.~\ref{fig:dif_polar} are included in the
fitting data base. However, comparing this result with the
prediction of Kaon-Maid demonstrates that the present work 
obeys the relation given by Eq.~(\ref{eq:polar_relation}) and 
provides a substantial improvement in the case of recoil
polarization observable $P$. 

As is pointed out in Ref.~\cite{biblap},
the relation between $P_\Lambda$ and $P_\Sigma$ given above seems to hold
only at higher $W$. Indeed, it is found that in certain kinematics 
region the relation fails to reproduce experimental data. 
In the present work, by comparing the solid curves in Fig.~\ref{fig:dif_polar}
and the result of my previous work (Fig.~6 of Ref.~\cite{mart_thr}), 
I find that this relation seems to work very well near the threshold region. 

In spite of the nice agreement between experimental data and the present
result, Fig.~\ref{fig:dif_polar} also indicates that 
more data at backward and very forward angles are desired to
support the present conclusion, especially that on the relation between
the $\Lambda$ and $\Sigma^0$ polarizations near threshold.  

The polarization of $\Sigma^+$ in the $\gamma+p\to {K^0} +\Sigma^+$ 
process has been also measured with a similar technique, because
 $\Sigma^+$ decays to $p\pi^0$ and $n\pi^+$ 
\cite{lawall}. However, since the energy of measurement has been 
averaged between threshold and $W=1.95$ GeV, the corresponding 
energy is obviously beyond the present interest.

\section{New Crystal Ball Data}
\label{sec:crystal-ball}

\begin{figure}[t]
  \begin{center}
    \leavevmode
    \epsfig{figure=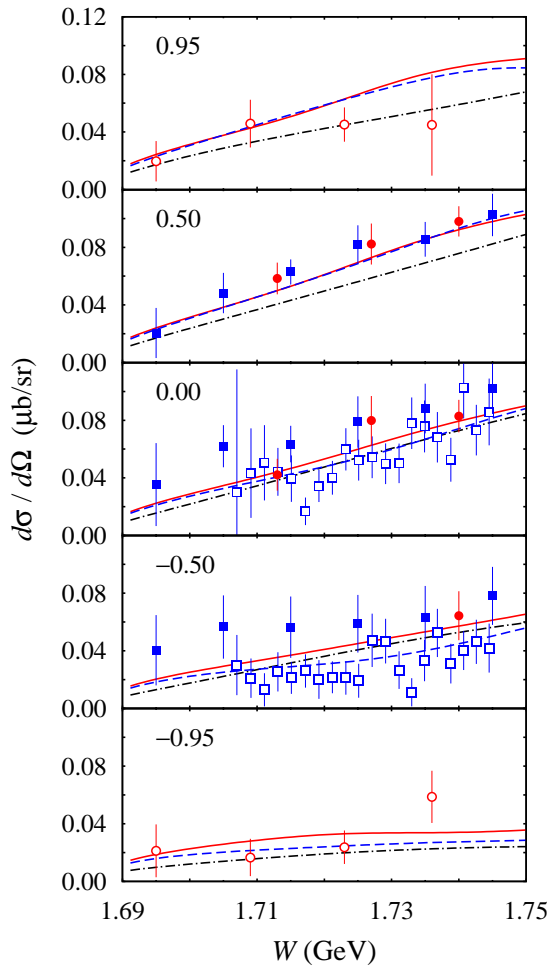,width=75mm}
    \caption{(Color online) Differential cross section
      for the $\gamma p\to K^+\Sigma^0$ channel. The effect
      of the new Crystal Ball data \cite{Jude:2013jzs} 
      is shown by the
      deviation of the dashed lines from the solid lines.
      Notation for the solid and dash-dotted lines as well as
      for experimental data points is given in the caption of 
      Fig. \ref{fig:dif_th}. The corresponding 
      value of $\cos\theta$ is given in each panel.}
   \label{fig:cb-data} 
  \end{center}
\end{figure}

\begin{table}[!t]
  \centering
  \caption{Comparison between the extracted resonance 
    photoproduction parameters obtained 
    from fits to experimental data with and without the new Crystal
    Ball data \cite{Jude:2013jzs} to those of the PDG estimate \cite{olive}.}
  \label{tab:resonance-CB}
  \begin{ruledtabular}
  \begin{tabular}[c]{lccc}
    Resonance parameters & Without & With & PDG \\
\hline
$N(1700)D_{13}$\\
\hline
$M_R$ (MeV)                          &1716 &1735 &$1700\pm 50$ \\
$\Gamma_R$ (MeV)                     &250  &100  &$150^{+100}_{-50}$ \\
$\beta_{K\Sigma}$                    &$3.0\times 10^{-2}$&$3.0\times 10^{-2}$&$\cdots$ \\
$\phi$ (deg)                         &$163$&153 &$\cdots$ \\
$A_{1/2}(p)$ ($10^{-3}$ GeV$^{-1/2}$)&$40$ &$40$ &$15\pm 25$ \\
$A_{3/2}(p)$ ($10^{-3}$ GeV$^{-1/2}$)&$-6$ &$-18$&$-15\pm 25$ \\
\hline
$\Delta(1700)D_{33}$\\
\hline
$M_R$ (MeV)                          &1692 &1693 &$1700^{+50}_{-30}$ \\
$\Gamma_R$ (MeV)                     &400  &200  &$300\pm 100$ \\
$\beta_{K\Sigma}$      &$6.4\times 10^{-5}$&$1.2\times 10^{-4}$&$\cdots$ \\
$\phi$ (deg)                         &$63$ &90   &$\cdots$ \\
$A_{1/2}(p)$ ($10^{-3}$ GeV$^{-1/2}$)&$170$&149  &$140\pm 30$ \\
$A_{3/2}(p)$ ($10^{-3}$ GeV$^{-1/2}$)&$110$ &110   &$140\pm 30$ \\
\hline
$N(1710)P_{11}$\\
\hline
$M_R$ (MeV)                          &1727    &1740 &$1710\pm 30$  \\
$\Gamma_R$ (MeV)                     & 50    &50 &$100^{+150}_{-50}$  \\
$\beta_{K\Sigma}$                    &$5.5\times 10^{-3}$ &$2.6\times 10^{-3}$ &$\cdots$ \\
$\phi$ (deg)                         &$40$    &42 &$\cdots$ \\
$A_{1/2}(p)$ ($10^{-3}$ GeV$^{-1/2}$)& 43    &57 &$40\pm 20$ \\
\hline
$N(1720)P_{13}$\\
\hline
$M_R$ (MeV)                          &1700 &1700 & $1720^{+30}_{-20}$\\
$\Gamma_R$ (MeV)                     & 150 &150 & $250^{+150}_{-100}$\\
$\beta_{K\Sigma}$      &$1.2\times 10^{-4}$&$4.2\times 10^{-4}$ &$\cdots$ \\
$\phi$ (deg)                         &360  &360 &$\cdots$ \\
$A_{1/2}(p)$ ($10^{-3}$ GeV$^{-1/2}$)& 120  &120 &$100\pm 20$ \\
$A_{3/2}(p)$ ($10^{-3}$ GeV$^{-1/2}$)& 120  &$120$ &$150\pm 30$ \\
\hline
$N$ & 331 & 560 & $\cdots$ \\
$\chi^2/N$ & 1.07 & 1.09& $\cdots$\\
  \end{tabular}
  \end{ruledtabular}
\end{table}

Recently, a new measurement of the $K^+\Sigma^0$ differential cross
section has been performed by the Crystal Ball collaboration 
at MAMI with finer energy bins \cite{Jude:2013jzs}. For the present work
the result of this measurement increases the number of
experimental data by 229 points. Since the number of data 
points in the database is almost doubled, the inclusion of
these new data might have a significant influence on the 
previous result. 
To this end, I have refitted my previous model by including
the new data. The relevant parameters obtained in
this case, i.e. the photoproduction parameters, 
are shown in Table \ref{tab:resonance-CB}, where the result
from the previous model (see Table \ref{tab:resonance}) 
along with the corresponding PDG estimate are also displayed 
for comparison. Obviously, there are no dramatic changes in 
the parameters. Furthermore in both fits (with and without 
the new Crystal Ball data) the fitted parameters are still 
consistent with the PDG values. Since the new measurement 
has been performed mostly in the  backward direction,
the corresponding effect is clearly more apparent in
this kinematics, as shown in Fig. \ref{fig:cb-data}.
Note that the new Crystal Ball data seem to be more consistent with
the SAPHIR data \cite{Glander:2003jw} and, consequently,
the result of including the new data at the 
corresponding kinematics
is lowering the predicted differential cross section, 
approaching the prediction of Kaon-Maid \cite{kaon-maid}.

\section{Result for Electroproduction}
\label{result-electroproduction}
Experimental data for low energy kaon electroproduction 
were unavailable until the CLAS
\cite{Ambrozewicz:2006zj} and MAMI A1 collaborations 
\cite{patrick_EPJA} published their recent measurements.
Note that in the old database 
the lowest energy available is 1.930 GeV \cite{old-data} and, 
therefore, they are irrelevant for the present study.
Since the discussion on MAMI data will be given in details in 
the next section, I will only focus on the CLAS data
in this section.

\begin{figure}[!t]
  \begin{center}
    \leavevmode
    \epsfig{figure=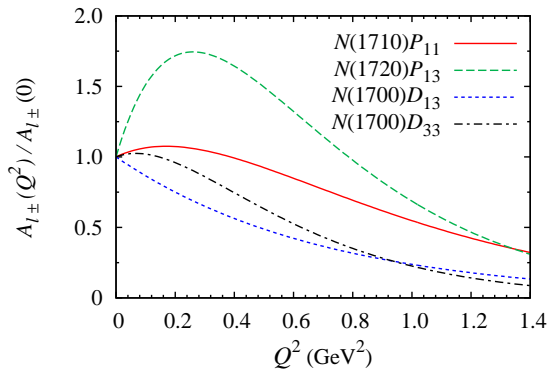,width=75mm}
    \caption{(Color online) $Q^2$ dependence of the resonance 
      multipoles for the nucleon resonances used in the present
      analysis.}
   \label{fig:ff_nstar} 
  \end{center}
\end{figure}

In the present work the $Q^2$ dependence of the resonance multipoles 
is given by \cite{Mart:2007mp}
\begin{eqnarray}
  \label{eq:Q2_dependence}
  A_{\it l\pm}(Q^2) = A_{\it l\pm}(0)\, (1+\alpha_{N^*} Q^2)\,
  e^{-\beta_{N^*} Q^2}~,
\end{eqnarray}
where $\alpha_{N^*}$ and $\beta_{N^*}$ are fitting parameters
given in Table~\ref{tab:resonance}. Note that this 
parameterization is used in Maid 
for the higher resonances  (see Eq.~(47) of Ref.~\cite{Drechsel:2007if}). 
The values of $\alpha_{N^*}$ and $\beta_{N^*}$ 
given in Table~\ref{tab:resonance} are certainly not comparable
to those of Maid, because in Maid the $A_{1/2}$ and $A_{3/2}$ 
amplitudes have different parameterization, whereas in the
present work they are the same. Nevertheless, the same trend
can be observed, e.g., in the case of $N(1720)P_{13}$, where
the value of  $\alpha_{N^*}$ tends to be large, while
the value of $\beta_{N^*}$ is moderately low. In general, except
for the $N(1700)D_{13}$ it is found that the value of  $\alpha_{N^*}$ 
is not zero, so that the amplitudes $A_{\it l\pm}(Q^2)$ increase from 
$A_{\it l\pm}(0)$ at low $Q^2$ and monotonically decrease at 
higher  $Q^2$ values. This behavior for the four nucleon resonances
used in the present analysis is exhibited in Fig.~\ref{fig:ff_nstar}.

\begin{figure}[!t]
  \begin{center}
    \leavevmode
    \epsfig{figure=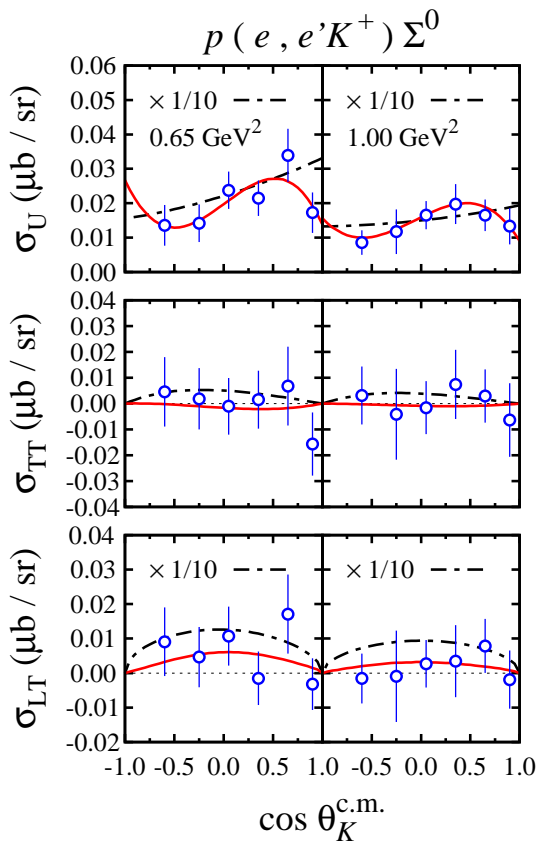,width=75mm}
    \caption{(Color online) Separated differential cross sections for 
      kaon electroproduction $e+p\to e'+K^++\Sigma^0$ as a function
      of kaon scattering angles at $W=1.725$ GeV and for two 
      different values of $Q^2$ (the values are shown in the top panels). 
      Experimental data are from the CLAS collaboration
      \cite{Ambrozewicz:2006zj}. 
      Notation of the curves is as in Fig.~\ref{fig:dif_th}.
      Note that $\sigma_i\equiv d\sigma_i/d\Omega$, where $i={\rm U,TT}$ and
      LT.
      Predictions of Kaon-Maid in the top and bottom panels have 
      been renormalized by a factor of 1/10 in order to fit on the scale.}
   \label{fig:ambroz} 
  \end{center}
\end{figure}

Unlike the old data, the latest CLAS data have been 
already separated in terms of the unpolarized 
differential cross section $\sigma_{\rm U}\equiv d\sigma_{\rm T}/d\Omega + 
\epsilon d\sigma_{\rm L}/d\Omega$, the transverse one 
$\sigma_{\rm TT} \equiv d\sigma_{\rm TT}/d\Omega$, and 
the longitudinal-transverse interference one 
$\sigma_{\rm LT} \equiv d\sigma_{\rm LT}/d\Omega$.
Comparison of these data with the results of present
work and Kaon-Maid is shown in Fig.~\ref{fig:ambroz}.
In contrast to the prediction of Kaon-Maid, the CLAS data are remarkably
much smaller, almost one order of magnitude.
The over prediction of Kaon-Maid in the finite $Q^2$ region will
be discussed in the next Section, when the result of the present
work is compared 
with the new MAMI data. The agreement of the result of the present work
with the CLAS data is clearly not surprising, because the data
are fitted. However, I would like to notice here that
in the case of unpolarized differential 
cross section (top panels) the cross
section shows a certain structure in the angular distribution, i.e.,
a peak at $\cos\theta\simeq 0.5$ and a tendency to increase at backward
direction. This indicates that the $t$ and $u$ channels should contain
a significant longitudinal coupling in the case of electroproduction,
which seems to disappear in the photoproduction case as displayed 
in Fig.~\ref{fig:dif_th}. Figure \ref{fig:ambroz} also exhibits that
the virtual photoproduction cross section is in fact very small, much
smaller than the predictions of isobar models as well as old data.
Nevertheless, this is consistent with the photoproduction cross 
section, provided that there is no dramatic increase of the 
cross section in the $Q^2$ distribution, which will be discussed
in the next section.

\begin{figure}[!t]
  \begin{center}
    \leavevmode
    \epsfig{figure=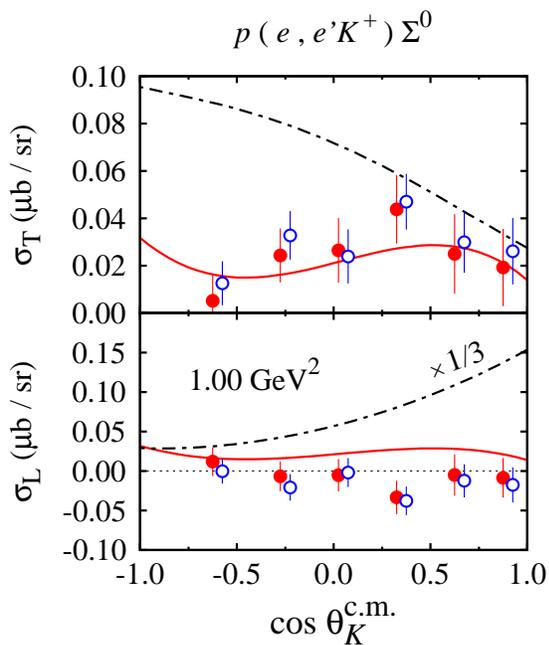,width=75mm}
    \caption{(Color online) As in Fig.~\ref{fig:ambroz}, but for 
      separated transverse (T) and longitudinal (L)
      differential cross sections at $W=1.75$ GeV and 
      $Q^2=1.0 $ GeV$^2$.
      Note that all experimental data shown in this figure 
      \cite{Ambrozewicz:2006zj} are not used
      in the fit. Solid and open circles 
      indicate two different methods of the 
      $\sigma_{\rm L}$$-$$\sigma_{\rm T}$ separation, which are
      slightly shifted for the sake of visibility.
      Further explanation 
      of the data can be found in
      Ref.~\cite{Ambrozewicz:2006zj} as well as in the 
      JLab Experiment CLAS Database http://clasweb.jlab.org/physicsdb/.
      In the lower panel, 
      prediction of Kaon-Maid has been renormalized 
      by a factor of 1/3 in order to fit on the scale.}
   \label{fig:ambroz_lt} 
  \end{center}
\end{figure}

Finally, it should also be noted that the 
longitudinal-transverse (LT) separation of
the cross section has been also performed in Ref.~\cite{Ambrozewicz:2006zj}.
The lowest energy available for this separation is 1.75 GeV, which is 
slightly beyond the upper limit of the present calculation. 
Nevertheless, for the sake of completeness and future 
$\sigma_{\rm L}$$-$$\sigma_{\rm T}$ separation technique 
the prediction of the present work along 
with the experimental data is exhibited 
in Fig.~\ref{fig:ambroz_lt}, where the prediction of Kaon-Maid is also
shown for comparison. In general, the prediction of the present work 
provides a fair agreement with experimental
data. Interestingly, Kaon-Maid predicts a large and forward peaking 
longitudinal cross section, whereas in the case of transverse cross 
section it shows very different behavior. 
It can obviously be seen that the shape of unpolarized 
cross section of both models  shown in the upper panel of 
Fig.~\ref{fig:ambroz} clearly originates from the separated ones 
as reflected by 
Fig.~\ref{fig:ambroz_lt}. It is also apparent that the result of 
the present calculation 
over estimates the data near forward angles. This happens presumably because
the corresponding energy is already beyond the upper limit of current
study. However, at this kinematics the 
longitudinal cross section data have negative values, which is 
certainly difficult to be reproduced by the model, since by
definition $\sigma_{\rm L}\propto |H_5|^2+|H_6|^2 \ge 0$, where $H_5$ and
$H_6$ are functions of the longitudinal CGLN amplitudes $F_5$ and
$F_6$ \cite{germar}.
Therefore, the present calculation recommends a new analysis on the
$\sigma_{\rm L}$$-$$\sigma_{\rm T}$ separation by imposing a new
constraint on the longitudinal cross section, i.e. $\sigma_{\rm L}\ge 0$.
Such a constraint could be expected to reduce some uncertainties in the separation
technique given in Ref.~\cite{Ambrozewicz:2006zj}.

\section{New MAMI Data at Low $Q^2$}
\label{sec:mami}
Recently, the A1 Collaboration at MAMI, Mainz, has measured the kaon 
electroproduction process $e+p\to e'+K^++\Sigma^0$ close to the production 
threshold and at very low virtual photon momentum transfers, 
i.e. $Q^2=0.030-0.055$ GeV$^2$ \cite{patrick_EPJA}. 
These new data is obviously of
interest, because they can be expected to shed new information
on the transition between photo- and electroproduction process.
This transition corresponds to the longitudinal coupling in the
process and therefore is very crucial for investigation of
the electromagnetic form factors, especially those of kaons
and hyperons for which no stable target exists.

\begin{figure}[!t]
  \begin{center}
    \leavevmode
    \epsfig{figure=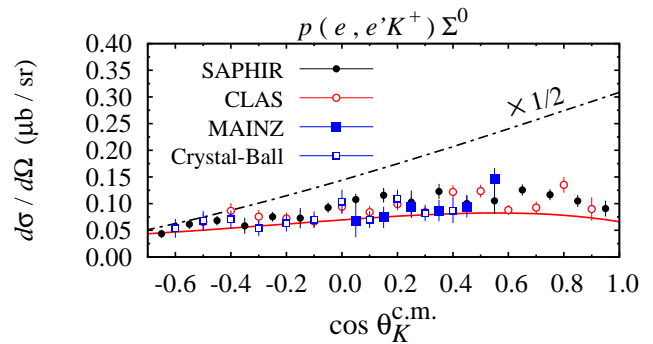,width=85mm}
    \caption{(Color online) Comparison between 
      new electroproduction data at 
      $W=1.75$ GeV and $Q^2=0.036$ GeV$^2$ from MAMI
      \cite{patrick_EPJA} 
      with photoproduction
      data from the SAPHIR \cite{Glander:2003jw}, CLAS 
      \cite{Bradford:2005pt}, and Crystal Ball \cite{Jude:2013jzs}
      collaborations, along with the prediction 
      of the present work (solid line) and Kaon-Maid
      (dash-dotted line). Prediction of Kaon-Maid has 
      been renormalized by a factor of 1/2 in order to fit on the scale. 
      Note that all data shown in this
      figure are not used in the fitting process since the corresponding 
      energies are higher than the upper energy limit. Furthermore, the 
      predicted differential cross sections are calculated for the
      electroproduction case ($Q^2=0.036$ GeV$^2$).}
   \label{fig:patrick} 
  \end{center}
\end{figure}

More than a decade ago, Niculescu {\it et al.} \cite{gabi98} have 
measured the $e+p\to e'+K^++\Lambda$ process and found that
the longitudinal cross section $d\sigma_L/d\Omega$ 
at $Q^2=0.52$ GeV$^2$ is significantly large, almost as large 
as the transverse one $d\sigma_T/d\Omega$. 
In order to reproduce these data an isobar model must 
dramatically increases both cross sections from photon point 
up to a certain value of $Q^2$ (see Fig. 5 of Ref. \cite{triest99}).
The revised analysis \cite{mohring_2003} of the same data 
found that the longitudinal
cross section is much smaller than the previous ones. 
Recent result from Jefferson Lab improves these data significantly
\cite{Ambrozewicz:2006zj}. However, in spite of
this substantial improvement, extrapolation of the combined cross
section $d\sigma_T/d\Omega+\epsilon d\sigma_L/d\Omega$ to the photon
point overshoots the experimental photoproduction data 
\cite{Ambrozewicz:2006zj}. 
Furthermore, due to the detector properties, 
there were no data points available in the range
of $Q^2=0.0-0.5$ GeV$^2$. Thus, the problem of dramatic increase
in the cross sections remained unsolved. 

\begin{figure}[!t]
  \begin{center}
    \leavevmode
    \epsfig{figure=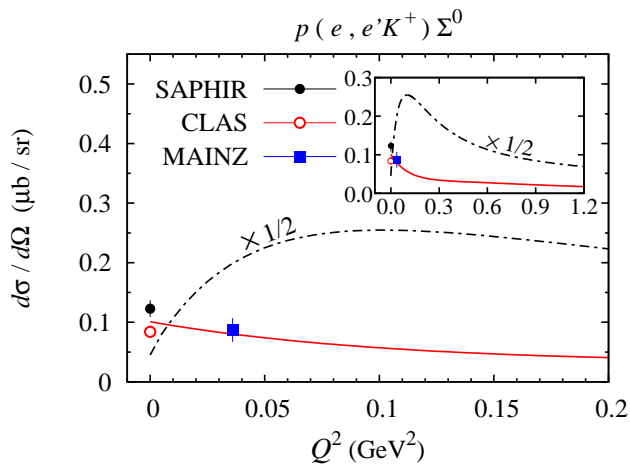,width=85mm}
    \caption{(Color online) As in Fig.~\ref{fig:patrick}
      but for the virtual photon momentum transfer $Q^2$ distribution.
    The insert shows comparison between Kaon-Maid with the 
    result of present work for higher $Q^2$ values.}
   \label{fig:patrick_q2} 
  \end{center}
\end{figure}

Although fitted to the different electroproduction data
\cite{old-data}, the $Q^2$ evolution of the cross section 
for the $e+p\to e'+K^++\Sigma^0$ channel of
Kaon-Maid exhibits the same behavior. 
However, this situation seems to be improved after the A1 Collaboration
published its result \cite{patrick_EPJA}. As shown in 
Figs.~\ref{fig:patrick} and \ref{fig:patrick_q2} the transition 
between photo- and electroproduction is found to be very smooth. 
In fact, within their error bars the data seem to be consistent.
Nevertheless, unlike the prediction of Kaon-Maid, the present 
calculation predicts an excellent agreement with the new MAMI data,
as shown clearly in Fig.~\ref{fig:patrick}.

Figure \ref{fig:patrick_q2} shows that the differential 
cross section just
monotonically falls off as the virtual photon momentum 
$Q^2$ increases from zero, in contrast
to the prediction of Kaon-Maid. The latter is understandable,
since it was fitted to the old data, which are scarce,
have large error bars and scattered in a wide range of 
kinematics \cite{old-data}. Thus, Fig. \ref{fig:patrick_q2}
indicates that for low energy case 
the longitudinal coupling in the $K^+\Sigma^0$
channel is relatively small and the shape of the 
electroproduction cross section
is practically driven by the conventional 
electromagnetic form factors. Obviously, more
electroproduction data with the same kinematics but
within the range of  
$Q^2=0.05-0.5$ GeV$^2$ are needed to support
the present conclusion. Such experimental data will be
available in the near future from MAMI collaboration
\cite{patrick-private}.

\section{Electromagnetic Form Factor of the Neutral Kaon}
\label{sec:form-factor}
In the previous paper \cite{mart_k0lambda} I have investigated the
effect of $K^0$ charge (electromagnetic) form factor on the
longitudinal differential cross section of the $e + n\to e'+K^0+\Lambda$
process. By using the the light-cone quark (LCQ) 
model~\cite{ito1} it is found that the form factor can raise the longitudinal 
cross section up to 50\%. In view of this promising result, 
it could be expected that experimental data with about 10\% 
error-bars would be able to experimentally prove the existence of 
this form factor in the process and, simultaneously, to select  
the appropriate $K^0$ form factor.

Given the large effect of $K^0$ charge form factor 
on the $K^0\Lambda$ longitudinal cross section,
it is clearly of interest to investigate the effect 
on both $K^0\Sigma^+$ and $K^0\Sigma^0$ channels
studied in the present work,
where the neutral kaon can directly interact with the virtual photon in
the $t$-channel. For this purpose I use the same form factor models
as in my previous study \cite{mart_k0lambda}, i.e., the light-cone 
quark (LCQ) model \cite{ito1} and the quark-meson vertex (QMV) 
model \cite{buck,buck1}. 

\begin{figure}[!]
  \begin{center}
    \leavevmode
    \epsfig{figure=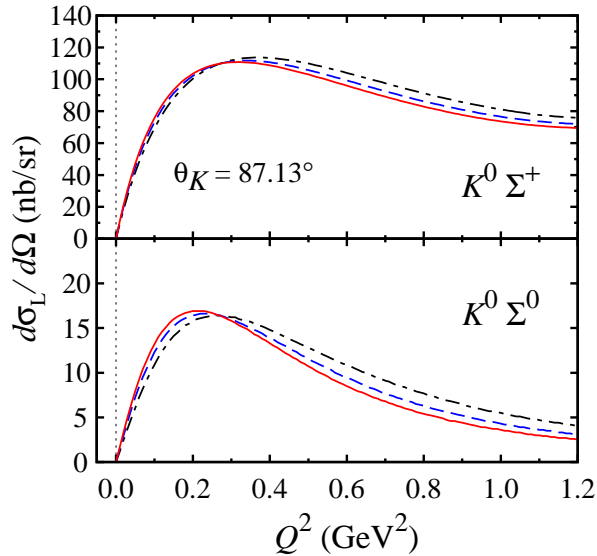,width=80mm}
    \caption{(Color online) Longitudinal differential cross section of the 
      neutral kaon electroproduction $e+p\to e'+K^0+\Sigma^+$ (a)
      and  $e+n\to e'+K^0+\Sigma^0$ (b), as a function
      of the virtual photon momentum squared $Q^2$ 
      at $W=1.72$ GeV and for kaon scattering angle $87.13^\circ$. 
      Solid lines show the calculation with a $K^0$ form 
      factor obtained in the LCQ model \cite{ito1} 
      while dashed lines are obtained by using the 
      QMV model \cite{buck,buck1}. The dash-dotted lines are obtained 
      from a computation with the $K^0$ pole excluded. }
   \label{fig:k0ff} 
  \end{center}
\end{figure}

The result for both $K^0\Sigma^+$ and $K^0\Sigma^0$ isospin channels
is shown in Fig. \ref{fig:k0ff}. Obviously, the effect found in both
cases is milder than that found in the $K^0\Lambda$ channel 
\cite{mart_k0lambda}, which is understandable, 
since the form factor in the present case is multiplied with the 
$g_{K\Sigma N}$ coupling constant. As shown in Table \ref{tab:background},
the value of the $g_{K\Sigma N}$ in the present case is 
about 60\% smaller than that of the $g_{K\Lambda N}$. Furthermore, 
I also note that in the case of $K^0\Lambda$ channel
contribution from the background terms is significantly larger
than that of the resonance terms
(see Fig.~1 of Ref. \cite{mart_k0lambda}). This is not the case
in  both $K^0\Sigma^+$ and $K^0\Sigma^0$ channels
(see Fig. \ref{fig:contrib}). Nevertheless, it is worth
to mention that in both channels there is moderate sensitivity to
these form factors in the range of 
$Q^2\approx 0.4 - 1.0$ GeV$^2$, where the largest one 
(up to about 10\%) originates from the LCQ model. The latter 
corroborates the finding of my previous
work \cite{mart_k0lambda}. 

From the above result I could conclude that in kaon electroproduction 
process only the $\gamma n\to K^0\Lambda$ channel seems to be the most promising
process for investigating 
the neutral kaon charge form factor. However, due to the
lack of free neutron target, the $\gamma p\to K^0\Sigma^+$ channel 
shown in the upper panel of Fig. \ref{fig:k0ff} could
become another alternative for this purpose, provided that more precise 
experimental measurements with about 5\% error bars along with a more accurate
$\sigma_{\rm L}$$-$$\sigma_{\rm T}$ separation technique 
were available.
Such a measurement is presumably suitable for future
experiment at MAMI in Mainz.


\section{Summary and Conclusions}
\label{sec:summary}
I have analyzed the elementary photo- and electroproduction 
of $K\Sigma$ for all four possible isospin channels near their 
production thresholds. To this end I have used an isobar
model based on suitable Feynman diagrams for the background
terms, for which all unknown
parameters such as hadronic coupling constants and electromagnetic
form factor cut offs were extracted from experimental data. 
For the resonance terms I used the Breit-Wigner form of multipoles, 
in which the values of photon couplings were taken from the PDG.
It is found that near the thresholds the four isospin channels of $K\Sigma$ 
photoproduction are mostly driven by their background terms, 
instead of the resonance terms as in the case of $K\Lambda$ photoproduction. 
Furthermore, in contrast to the $K^+\Lambda$ channel, 
the present study indicates that the hyperon resonances 
do not play an important role in $K\Sigma$ channels.

Whereas the result of the present calculation for the proton channels 
provides a nice agreement with experimental data as well as a substantial 
improvement of my previous work, the prediction of the present analysis 
for the neutron channels 
are plagued with large uncertainties that originate from the uncertainties 
in the values of helicity photon couplings given by the PDG. 
The present study also proves the validity of the 
relation between $\Lambda$ and $\Sigma^0$ polarizations, i.e.,
$P_\Lambda=-(1/3)P_\Sigma$, at energies near thresholds. 

The extracted longitudinal differential cross section from the CLAS experiment
is found to be too small. This finding suggests the necessity for a new
extraction method that imposes the condition that the cross section
values are always positive.
The new MAMI electroproduction data at very low $Q^2$ can
be nicely reproduced, although they were not included
in the present analysis. These new data support the smooth transition 
behavior from photo- to electroproduction, which is 
exhibited by the present work, but not Kaon-Maid.
Therefore, the large longitudinal term coming from the $D_{13}(1895)$
resonance in Kaon-Maid is not proved. 

Finally, the effect of neutral kaon charge form factor on the 
longitudinal cross sections of $K^0\Sigma^+$ and $K^0\Sigma^0$
channels is found 
to be smaller than that obtained in the $K^0\Lambda$ channel.
Nevertheless, the $K^0\Sigma^+$ channel could become an alternative
process for investigation of this form factor, provided 
that the corresponding
longitudinal cross section can be accurately extracted.

\section*{Acknowledgment}
The author thanks Tom Jude and Daniel Watts for providing him with
the new Crystal Ball $K\Sigma$ data. Useful discussion with
Igor I. Strakovsky and Reinhard Schumacher is gratefully acknowledged.
This work has been partly supported by the Research-Cluster-Grant-Program 
of the University of Indonesia, under contract No. 1709/H2.R12/HKP.05.00/2014.

\end{document}